\documentclass{agujournal2018}

\usepackage[utf8]{inputenc}
\usepackage[english]{babel}
\usepackage{amsmath}
\usepackage{amsthm}
\usepackage{amsfonts}
\usepackage{siunitx}

\journalname{JGR: Solid Earth}

\begin{document}

\title{Self-similar distributions of fluid
  velocity and stress heterogeneity in a dissolving porous
  limestone}

\authors{Gaute Linga\affil{1}, Joachim Mathiesen\affil{1}, and François Renard\affil{1,2,3}}
\affiliation{1}{Niels Bohr Institute, University of Copenhagen, Blegdamsvej 17, DK-2100 Copenhagen, Denmark.}
\affiliation{2}{Dept.~of Geosciences, Physics of Geological Processes, University of Oslo, Norway.}
\affiliation{3}{Université Grenoble Alpes, ISTerre, CS40700, Grenoble, France.}

\vspace{1cm}
Manuscript published in J.\ Geophys.Res.\ Solid Earth {\bf 122}, 1726–1743 (2017), 

{\bf doi}:10.1002/2016JB013536.

\begin{keypoints}
\item The fluid-solid interaction problem is solved numerically in a
  porous limestone core sample at several stages of dissolution.
\item Probability density functions of fluid velocity and solid stress
  evolve self-similarly as porosity and flow rate is varied.
\item The unified distributions provide an additional explanation of
  the sensitivity of rocks to failure due to fluid flow and
  dissolution.
\end{keypoints}

\begin{abstract}
  In a porous rock, the spatial distribution of the pore space induces
  a strong heterogeneity in fluid flow rates and in the stress
  distribution in the rock mass. If the rock microstructure evolves
  through time, for example by dissolution, fluid flow and stress will
  evolve accordingly. Here, we consider a core sample of porous
  limestone that has undergone several steps of dissolution. Based on
  3D X-ray tomography scans, we calculate numerically the coupled
  system of fluid flow in the pore space and stress in the solid. We
  determine how the flow field affects the stress distribution both at
  the pore wall surface and in the bulk of the solid matrix. We show
  that, during dissolution, the heterogeneous stress evolves in a
  self-similar manner as the porosity is increased. Conversely, the
  fluid velocity shows a stretched exponential distribution. The
  scalings of these common master distributions offer a unified
  description of the porosity evolution, pore flow, and the
  heterogeneity in stress for a rock with evolving
  microstructure. Moreover, the probability density functions of
  stress invariants (mechanical pressure or von Mises stress) display
  heavy tails towards large stresses. If these results can be extended
  to other kinds of rocks, they provide an additional explanation of
  the sensitivity to failure of porous rocks under slight
  changes of stress.
\end{abstract}

\section{Introduction}
\label{sec:introduction}
Reactive fluid flow in porous rocks under stress is ubiquitous both in
nature and in industrial applications. Porous flow controls rock
weathering, diagenesis in the crust, karst formation, and large scale
fluid circulations at the origin of ore deposits\
\citep{jamtveit2012,bjorlykke1997}. Fluid flow coupled to deformation
of porous rocks control the degree to which earthquake-induced
deformation can drive transient or permanent changes in crustal
permeability~\citep{rice1976}. In fault zones, fluid may exert a pore
pressure large enough to reduce the apparent strength along the slip
surface, providing an explanation for the apparent low heat frictional
force observed on the San Andreas fault in California\
\citep{byerlee1990}. When coupled to rock transformations in fault
zones, this mechanism was also proposed to explain how long term
variations of fluid pressure could control the seismic cycle
\citep{sibson1992,gratier2003}.

Industrial applications include enhanced oil recovery, carbon dioxide
sequestration, hydraulic fracturing, and cement ageing. Injection of
CO$_2${} into geological formations, aquifers or depleted petroleum
reservoirs, poses a promising route to reduce greenhouse gas emissions
in the framework of Carbon Capture and Storage
\citep{iea2014energy}. Since such formations often contain carbonate
minerals, they may react with the injected CO$_2${} rich fluid,
resulting in changes of the pore-space geometry that couples to
deformation \citep{rohmer2016}. This modifies both reactive surface
area, porosity, permeability and, finally, the ability of the rock to
store carbon in minerals\ \citep{noiriel2004,noiriel2005}.

Search for macroscopic properties, such as porosity or permeability,
from local scale description of microstructures, show that the
presence of heterogeneities controls nonlinearities in the transport
properties of a porous medium \citep{bernabe1995}. The recent
development of the field of Digital Rock Physics now allows to
calculate various mechanical and transport properties in rocks based
on the full 3D images of the samples measured by X-ray microtomography
\citep{arns2002,andra2013,oren2007}. Fluid flow at the pore scale has
been studied using 3D porous media extracted by X-ray microtomography
to characterize processes such as capillary trapping, CO$_2${}
sequestration, multiphase flow, solute transport, time-dependent
evolution of microstructures during fluid-rock interactions
\citep{blunt2013,noiriel2015,bultreys2016,misztal2015detailed}, and
numerical modeling methods are reviewed in~\citep{meakin2009}. Elastic
properties change during rock transformation (dissolution and
precipitation) and depend on the initial
microstructure~\citep{wojtacki2015}. For example, if either micro- or
macropores dissolve preferentially in
a limestone rock, the resulting change of elastic parameters and
seismic wave velocities would be different~\citep{arson2015}.

On the one hand, simulations of flow through porous solids have
determined that there exists orders of magnitude variations in local
fluid velocity, even at the millimeter
scale~\citep{brown1987,bijeljic2013,deanna2013,leborgne2013},
indicating that both local pressure gradients and channeling flow are
important \citep{brown1987}. On the other hand, the study of coupled
fluid flow and solid deformation is the basis of the theory of
poro-elasticity~\citep{rice1976,coussy2004}. Here, we study the
coupling between stress and fluid flow in a porous rock that
dissolves. The complexity of fluid flow and stress
heterogeneity stems from randomness of the medium and the
possible coupling between forces in the solid and forces exerted by
the flowing fluid. The stress distribution in the solid phase, and in
particular at the solid-fluid interface, is highly heterogeneous at
the scale of grains and pores in the rock. Regions of high stress are
prone to stress-enhanced dissolution and crack formation, while
regions of low stress are prone to precipitation due to solute
transport in the pore space. These processes, over time, alter the
pore space geometry and constitute a feedback loop between flow and
deformation.

We characterize numerically how heterogeneities in stress, fluid flow,
and microstructures impact the hydro-mechanical behavior of a
limestone rock that has undergone several steps of dissolution. We aim
to answer the following questions:
\begin{enumerate}
\item How does single-phase fluid flow through the pore space of a
 rock sample under external load affect the stress distribution in
 it, and, in particular, what is the effect of a heterogeneous
 microstructure on the stress distribution?
\item How does rock dissolution modify the state of stress and yield
 strength in the solid?
\item How does dissolution in the rock modify single phase fluid flow?
\end{enumerate}
The main objectives are \emph{(1)} to examine whether the stress
distributions in the bulk of the solid and at the solid-fluid
interface can be described by a common probability density function
(PDF),  and \emph{(2)} to quantify the effects of dissolution,
i.e.~changes in the complex pore space, on the stress distribution and
flow properties. We address these objectives by computational means,
using the finite element method to solve the coupled fluid-solid
mechanics problem in three-dimensional digitized porous rocks. From
our computations, we achieve the fluid velocity field and the stress
field in both the fluid and in the solid. We further estimate the
mechanical pressure and von Mises stress in the solid under various
conditions of external and internal loading. We apply our method to a
sample of limestone that has undergone successive steps of dissolution
through the percolation of an acidic fluid. This sample was imaged in
3D before percolation and at three successive steps of dissolution
using synchrotron X-ray microtomography
\citep{noiriel2004,noiriel2005}.

The results of the present study can be of primary interest in domains
where the heterogeneous and multiscale nature of rocks plays a key
role, including, for example, oil and gas reservoir engineering,
CO$_2${} geological sequestration, and fracture mechanics.

\section{Model and method}
\label{sec:method}
In this section, we present the numerical methods and the computational
model used to calculate the state of stress in a porous solid with a
percolating fluid. As we are interested in the instantaneous effects
of steady fluid flow, we assume a timescale where the effect of
chemical reactions is negligible and where the pore space geometry
does not change---i.e.\ there is no evolution of the
microstructure. In this regime, the computational problem involves a
one-way coupling of normal stress from the fluid flow to the solid
stress field.

\subsection{Fluid flow in the pore space}
\label{sec:fluideqs}
The Navier-Stokes equations, governing the incompressible fluid flow
in the pores, are given by
\begin{linenomath*}
\begin{gather}
 \rho \left( \frac{\partial {\mathbf v}}{\partial t}  + ({\mathbf v} \cdot {\ensuremath{\mbox{\boldmath$ \nabla $}}{}} ) {\mathbf v} \right) - \mu {\ensuremath{\mbox{\boldmath$ \nabla $}}{}}^2 {\mathbf v} =  - {\ensuremath{\mbox{\boldmath$ \nabla $}}{}} {P},
 \label{eq:ins_mom}\\ 
 \ensuremath{\mbox{\boldmath$ \nabla $}} \cdot \mathbf v  = 0, \label{eq:ins_mass} 
\end{gather}
\end{linenomath*}
defined on a domain $\Omega_\ell$. Here, $\mathbf v ( \mathbf x, t)$ is the
velocity field, $P$ is the pressure of the fluid, $\rho$ is the
(constant) fluid density, and $\mu$ is the dynamic viscosity.  Closure
is obtained by supplying an initial condition
$\mathbf v (\mathbf x, 0) = \mathbf v_0 (\mathbf x )$, and a set of boundary conditions:
\begin{linenomath*}
\begin{eqnarray}
 {\mathbf v} ({\mathbf x}, t) &=& \mathbf 0 \quad \textrm{for} \quad \mathbf x \in \Gamma_\textrm{wall}, \\
 {P} ({\mathbf x}, t) &=& P_\textrm{in} \quad \textrm{for} \quad \mathbf x \in \Gamma_\textrm{in}, \\
 {P} ({\mathbf x}, t) &=& P_\textrm{out} \quad \textrm{for} \quad \mathbf x \in \Gamma_\textrm{out} .
\end{eqnarray}
\end{linenomath*}
Here,
$\Gamma = \Gamma_\textrm{wall} \cup \Gamma_\textrm{in} \cup \Gamma_\textrm{out}$
represents the entire boundary of $\Omega_\ell$, which we, for
now, assume does not deform in time, while $P_\textrm{in}$ and $P_\textrm{out}$ are
constant fluid pressures imposed at the inlet and outlet of the
system, respectively.

For a porous rock we consider that the characteristic length scale
$\ell_p$ of the pore space is small, such that the ratio between
inertial and viscous forces is low, i.e.~the Reynolds number
$\mathrm{Re} = \rho|{\mathbf v}| \ell_{\textrm{p}} /
\mu \ll 1$.
Thus the advection part of Eq.~\eqref{eq:ins_mom},
$ \partial {{\mathbf v}} / \partial {t} + ( {\mathbf v \cdot \ensuremath{\mbox{\boldmath$ \nabla $}}{}} ) {\mathbf v} $,
is assumed to be negligible. This assumption is verified, as in
experiments by \citet{noiriel2004} {(see
  Sec.~\ref{sec:geometry})}, the speed
$|{\mathbf v}|$ is in the range
1--$\SI{4e-3}{\meter\per\second}$, the typical pore size
$\ell_{\textrm{p}}$ is in the range 1--$\SI{3e-4}{\meter}$, and {the kinematic}
viscosity of water $\mu/\rho = \SI{1e-5}{\meter\squared\per\second}$, which
gives a Reynolds number $\mathrm{Re} < 0.1$.

In the limit of low Reynolds number, the Navier--Stokes equations
\eqref{eq:ins_mom} and \eqref{eq:ins_mass} reduce to the Stokes
equations, which are linear in velocity and pressure, and
can therefore be solved using optimized
linear solvers. The time dependence has now vanished,
and we are seeking the steady flow field. By introducing the
  dimensionless variables $\tilde{\mathbf x}, \tilde{\mathbf v}$ and $\tilde{P}$, implicitly defined
  by
\begin{linenomath*}
\begin{equation}
 {\mathbf x} = L \tilde{\mathbf x}, \quad
 {\mathbf v} = \frac{L}{{\mu}} (P_\textrm{in} - P_\textrm{out}) \tilde{\mathbf v}, \quad
 {P} {=} (P_\textrm{in} - P_\textrm{out}) \tilde{P} + \frac{P_\textrm{in} + P_\textrm{out}}{2},
 \label{eq:scalefluidvars}
\end{equation}
\end{linenomath*}
where $L$ is the system length, we obtain from equation
\eqref{eq:ins_mom} the well-known Stokes equation in non-dimensional
form,
\begin{linenomath*}
\begin{eqnarray}
 \tilde{\ensuremath{\mbox{\boldmath$ \nabla $}}{}}^2 \tilde{\mathbf v} &=& \tilde{\ensuremath{\mbox{\boldmath$ \nabla $}}{}}{P}, \label{eq:stokes_mom}\\
 \tilde{\ensuremath{\mbox{\boldmath$ \nabla $}}{}}\cdot \tilde{\mathbf v} &=& 0, \label{eq:stokes_mass}\\
 \tilde{\mathbf v} (\tilde{\mathbf x}) = \mathbf 0 &\textrm{for}& \tilde{\mathbf x} \in \tilde{\Gamma}_\textrm{wall}, \label{eq:stokes_bc_noslip}\\
 \tilde{P} ( \tilde{\mathbf x} ) = \frac 1 2 &\textrm{for}& \tilde{\mathbf x} \in \tilde{\Gamma}_\textrm{in}, \label{eq:stokes_bc_pin}\\
 \tilde{P} ( \tilde{\mathbf x} ) = - \frac 1 2 &\textrm{for}& \tilde{\mathbf x} \in \tilde{\Gamma}_\textrm{out}. \label{eq:stokes_bc_pout}
\end{eqnarray}
\end{linenomath*}
{Here, $\tilde{\ensuremath{\mbox{\boldmath$ \nabla $}}{}} = L \ensuremath{\mbox{\boldmath$ \nabla $}}{}$ is the scaled del operator, and $\tilde{\Gamma}$ (with the respective subscripts) is the scaled domain.}
Since these expressions are all independent of the constants
{$\rho$, }$\mu$, $P_\textrm{in}$, {and} $P_\textrm{out}$, all solutions to the Stokes equations are the same
up to a scaling constant and a shift in pressure.

The stress tensor in Stokes flow in dimensional quantities is given by
\begin{linenomath*}
\begin{equation}
  {\ensuremath{\mbox{\boldmath $ \sigma$}}} = - {P} \mathbf I + \mu ( {\ensuremath{\mbox{\boldmath$ \nabla $}}{}} {\mathbf v} + {\ensuremath{\mbox{\boldmath$ \nabla $}}{}} {\mathbf v}^\top ),
\end{equation}
\end{linenomath*}
which means that the dimensional strain tensor can be found from the
non-dimensional one{,}
\begin{equation}
  {\tilde{\ensuremath{\mbox{\boldmath $ \sigma$}}} = - \tilde{P} \mathbf I + \tilde{\ensuremath{\mbox{\boldmath$ \nabla $}}{}}\tilde{\mathbf v} + \tilde{\ensuremath{\mbox{\boldmath$ \nabla $}}{}} \tilde{\mathbf v}^\top,}
\end{equation}
by the transformation
\begin{linenomath*}
\begin{equation}
  \ensuremath{\mbox{\boldmath $\sigma$}} = (P_\textrm{in} - P_\textrm{out}) \tilde{\ensuremath{\mbox{\boldmath $\sigma$}}} - \frac{P_\textrm{in} + P_\textrm{out}}{2} \mathbf I.
  \label{eq:unscalefluidstress}
\end{equation}
\end{linenomath*}
As a consequence, for a given pore space geometry, performing \emph{one
 single} steady-state simulation is sufficient to obtain the stress
field for arbitrary inlet and outlet fluid pressures. Only the linear
transformation described above is required to achieve the field
resulting from the sought inlet/outlet conditions.

In the forthcoming, we use the following definitions:
\begin{linenomath*}
\begin{eqnarray}
 P_0 &=& \frac{P_\textrm{in} + P_\textrm{out}}{2} \qquad \textrm{(base pressure)} \\
 \Delta P &= &P_\textrm{in}-P_\textrm{out} \qquad \textrm{(pressure drop)}
\end{eqnarray}
\end{linenomath*}
to quantify the effect of fluid flow in the pore space.

\subsection{Fluid-solid stress coupling at the pore scale}
\label{subsec:fluidsolidcoupling}
At the boundary between fluid and solid, $\Gamma_\textrm{wall}$, the normal
stress should be continuous if the solid--fluid interfacial tension is
neglected:
\begin{linenomath*}
  \begin{equation}
    \label{eq:fluidsolidcoupling}
\left[\left[ { \ensuremath{\mbox{\boldmath $ \sigma$}} } \right] \right] = \ensuremath{\mbox{\boldmath $ \sigma$}} \cdot \mathbf n \big|_{\Gamma_\textrm{wall}^{(\textrm{s})}}
 - \ensuremath{\mbox{\boldmath $ \sigma$}} \cdot \mathbf n \big|_{\Gamma_\textrm{wall}^{(\ell)}} = \mathbf 0,
\end{equation}
\end{linenomath*}
where $\mathbf n$ is the unit normal at the interface, pointing into the
solid. The superscripts $(\textrm{s})$ and $(\ell)$ denote evaluation at the
solid and liquid sides of the interface, respectively. If we assume
that we consider time scales in which the solid does not deform due to
fluid flow
(Eqs.~\eqref{eq:stokes_mom}--\eqref{eq:stokes_bc_pout}),
the no-slip boundary condition \eqref{eq:stokes_bc_noslip} on the
fluid is valid, and hence the viscous stress boundary condition on the
solid is prescribed by the fluid. As mentioned above, this yields a
one-way coupling from the fluid to the solid phase which encompasses
computational simplification.

\subsection{State of stress in the solid phase}
For small deformations, the solid phase is described by linear
elasticity, such that stress, $\ensuremath{\mbox{\boldmath $ \sigma$}}$, and strain, $\ensuremath{\mbox{\boldmath $ \epsilon $}}$, are
related via Hooke's law,
\begin{linenomath*}
\begin{equation}
 \ensuremath{\mbox{\boldmath $ \sigma$}} = \frac{E}{1+\nu} \left[ \ensuremath{\mbox{\boldmath $ \epsilon $}} + \frac{\nu}{1-2\nu} \mathbf I \mathrm{tr} {(} \ensuremath{\mbox{\boldmath $ \epsilon $}} {)} \right],
\end{equation}
\end{linenomath*}
where $E$ is Young's modulus and $\nu$ is Poisson's ratio. The strain
tensor in the solid is given by
\begin{linenomath*}
\begin{equation}
 \ensuremath{\mbox{\boldmath $ \epsilon $}} = \frac{1}{2} \left( \ensuremath{\mbox{\boldmath$ \nabla $}} \mathbf u + \ensuremath{\mbox{\boldmath$ \nabla $}} \mathbf u^\top \right),
\end{equation}
\end{linenomath*}
where $\mathbf u (\mathbf x)$ is the displacement field. By considering the
static elastic field (i.e.~time scales much larger than the time
elastic waves take to propagate through the system), stress
equilibrium in the rock is expressed by
\begin{linenomath*}
\begin{equation}
 \ensuremath{\mbox{\boldmath$ \nabla $}} \cdot \ensuremath{\mbox{\boldmath $ \sigma$}} [ \mathbf u (\mathbf x) ] = \mathbf 0,
\end{equation}
\end{linenomath*}
where the right hand side is equal to zero since we neglect body
forces, such as gravity.

Closure of the equation system is obtained by supplying the following
boundary conditions. Inside the rock, at the pore-solid interface,
this boundary condition is given by Eq.\
\eqref{eq:fluidsolidcoupling}. At the \emph{outside} boundary of
the solid, $\Gamma_\mathrm{ext}$, i.e.~the part of the boundary which is not in
contact with the fluid, a prescribed normal traction (equivalent to a
pressure force) is imposed:
\begin{linenomath*}
\begin{equation}
 \ensuremath{\mbox{\boldmath $ \sigma$}} \cdot \mathbf n = - P_\mathrm{ext} \mathbf n, \quad \textrm{for} \quad \mathbf x \in \Gamma_\mathrm{ext}\setminus\Gamma_\textrm{bot},
\end{equation}
\end{linenomath*}
except at the bottom plane $\Gamma_\textrm{bot}$, where we apply a no-slip
condition on the displacement field,
\begin{linenomath*}
\begin{equation}
 \mathbf u (\mathbf x) = \mathbf 0, \quad \textrm{for} \quad \mathbf x \in \Gamma_\textrm{bot},
\end{equation}
\end{linenomath*}
in order to remove translational and rotational freedom and
thereby achieve uniqueness of solution. Doing so,
the force exerted by the fluid flowing in the pore space to the solid
surfaces are integrated as a boundary condition and therefore coupled
to the state of stress of the solid.

\subsection{Geometry and mesh of the porous samples}\label{sec:geometry}
\begin{figure}[htb]
 \centering
 \includegraphics[width=0.8\textwidth]{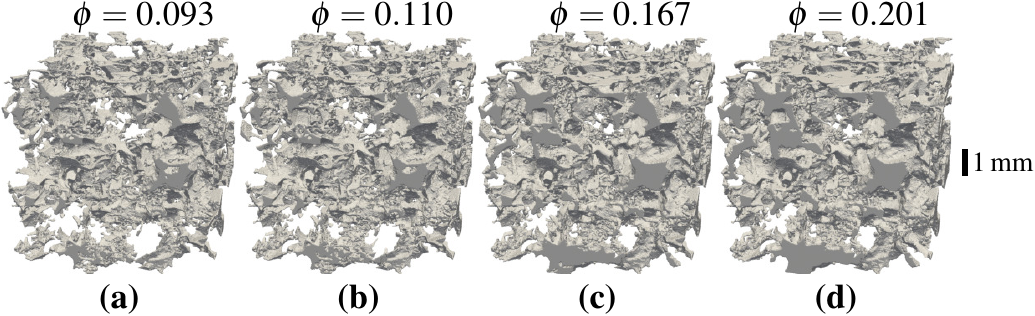}
 \caption{\label{fig:mesh_fluid}Fluid meshes of the limestone sample at
 different steps of dissolution. Sub-figures show \textbf{(a)}
 initial geometry, and \textbf{(b)} 1 step, \textbf{(c)} 2 steps
 and \textbf{(d)} 3 steps of dissolution. Fluid flow was from
 bottom to top during the experiments.}
\end{figure}

\begin{figure}[htb]
 \centering
 \includegraphics[width=0.88\textwidth]{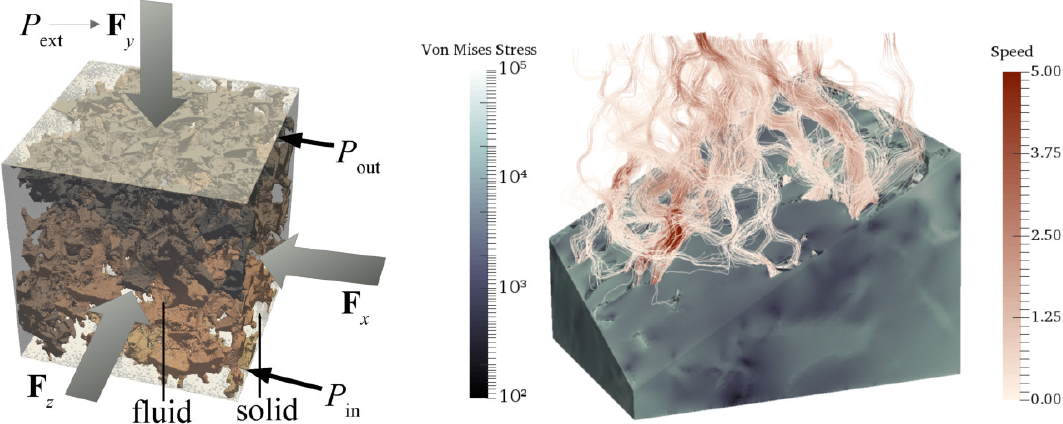}
 \caption{\label{fig:schematic}Left: Schematic set-up of the
   model. Right: Simulated 3D volume, after 3 dissolution steps, with
   fluid velocity streamlines and von Mises stress in the solid. The
   upper half has been clipped to display the fluid phase.}
\end{figure}

We consider a digital 3D rock sample which was studied and described
by \citet{noiriel2004}. It is a crinoïdal limestone of middle
Oxfordian age extracted from the Lérouville formation (Paris
Basin). Acidic fluid was injected into this sample, leading to
dissolution and porosity increase.  {The experiments were
  performed in the diffusion-controlled regime, at low injection rates
  to avoid dissolution fingering instabilites.}  The sample has
undergone three steps of dissolution and, between each dissolution
step, it was scanned in 3D, using X-ray microtomography at the
European Synchrotron Radiation Facility, at a voxel resolution of
$\SI{4.91}{\micro\meter}$. The
results are four digitized volumes: the initial sample before
percolation, and three volumes after the three stages of dissolution
(Fig.~\ref{fig:mesh_fluid}). The original 3D digitized volumes were
segmented to separate the pore space from the solid phase and
re-sampled at
$\SI{9.8}{\micro\meter}$ voxel
size. The volumes used in the present study have dimensions of $340^3$
voxels. The segmented images were prepared such that they constituted
one connected cluster both for the solid and fluid phases; i.e.~all
disconnected ``islands'' were removed. The removed disconnected pores
represented a fraction less than 0.05 of the total pore volume.  The
segmented volumes were then converted to a tetrahedral mesh for the
fluid phase using \textsc{iso2mesh} \citep{fang2009}, a
\textsc{MATLAB} interface to \textsc{TetGen} \citep{si2015} for the
surface mesh, and \textsc{CGAL} \citep{cgal2016} for the volumetric
mesh. The triangulated surface of this mesh is used as the inner
surface of the solid mesh. This surface mesh was then embedded into a
cubic surface mesh, which constituted the outer mesh (Figure
\ref{fig:schematic}). This cubic mesh was chosen to be slightly (about
2 \%) larger than the fluid mesh, such that the whole sample could be
loaded uniformly, yielding the same total force on each side of the
cube. A tetrahedral volume mesh was then generated between these
surfaces. In this way, \emph{(1)} the solid matrix can be loaded with
a uniform normal stress at the outside boundary, and \emph{(2)}
no-slip conditions can be appropriately applied for the fluid phase at
the entire surface $\Gamma$, except inlet $\Gamma_\textrm{in}$ and outlet
$\Gamma_\textrm{out}$.

\subsection{Computational model}
The rock we consider is a cubic, \emph{sealed}, elastic porous sample
which can be mechanically loaded along all axes, and saturated with a
steadily flowing single phase liquid. With this model, by varying the
flow rate and the externally applied stress, one may obtain \emph{(1)}
the fluid velocity field in the pore-space of the sample at the
different steps of dissolution; \emph{(2)} the stress distribution in
the solid space of the sample as a function of applied fluid pressure
and external stress; and \emph{(3)} the probability distributions of
invariants of the stress tensor throughout the sample surface or
volume.

\subsection{Implementation}
The coupled fluid-solid problem is solved numerically using the
\textsc{FEniCS/DOLFIN} framework
\citep{logg2012automated,logg2012dolfin}. The \textsc{FEniCS} project
is a collection of software for automated solution of differential
equations using the Finite Element Method (FEM), whereas DOLFIN is a
\textsc{C++/Python} library functioning as the main user interface to
\textsc{FEniCS}. It allows for efficient solution of differential
equations requiring only a weak (variational) formulation of the
problem to be specified.

\subsubsection{Fluid phase}\label{subsec:fluidphaseweak}
The fluid equations \eqref{eq:stokes_mom} to \eqref{eq:stokes_bc_pout}
are solved using a continuous-Galerkin method with first-order
Lagrange (P$_1$) elements both for the velocity and pressure fields
\citep{langtangen2002}. Since this mixed-space formulation of the
Stokes equations causes stability problems, as it violates the
Babuska--Brezzi condition \citep{brenner2008}, we use a pressure
stabilization technique. This amounts to allowing a small
grid-dependent compressibility which will smooth out the pressure
field solution \citep{langtangen2002}, i.e.
\begin{linenomath*}
\begin{equation}
 \ensuremath{\mbox{\boldmath$ \nabla $}} \cdot \mathbf v = \delta h^2 \ensuremath{\mbox{\boldmath$ \nabla $}}^2 P,
\end{equation}
\end{linenomath*}
where $h$ is the element size and $\delta$ is a heuristically chosen
parameter. {Here, we have omitted the tildes used for scaled
  units for the sake of visual clarity.} We verified that $\delta$
($= 0.04$) was chosen small enough for the absolute difference in
inlet/outlet flux to be well below $\SI{2}{\percent}$, such that the
mass of fluid is almost conserved.

The weak formulation of the fluid equations can thus be stated as the
following: Find $\mathbf v \in \mathcal{V}, P \in \mathcal{P}$ such
that
\begin{linenomath*}
\begin{multline}
 \int_{\Omega_\ell} \left( \ensuremath{\mbox{\boldmath$ \nabla $}} \mathbf v : \ensuremath{\mbox{\boldmath$ \nabla $}} \mathbf v' 
 - P \ensuremath{\mbox{\boldmath$ \nabla $}} \cdot \mathbf v' + P' \ensuremath{\mbox{\boldmath$ \nabla $}} \cdot \mathbf v 
 + \delta h^2 \ensuremath{\mbox{\boldmath$ \nabla $}} P \cdot \ensuremath{\mbox{\boldmath$ \nabla $}} P'
 \right) \mathrm{d} V \\
= -\int_{\Gamma_\textrm{in}} P_\textrm{in} \mathbf n \cdot \mathbf v' \mathrm{d} S -\int_{\Gamma_\textrm{out}} P_\textrm{out} \mathbf n \cdot \mathbf v' \mathrm{d} S
 \label{eq:weak_stokes}
\end{multline}
\end{linenomath*}
for all $\mathbf v' \in \mathcal{V}, P' \in \mathcal{P}$, and
$ \mathbf v (\mathbf x) = \mathbf 0$ for $\mathbf x \in \Gamma_\textrm{wall}$. Here, $\mathcal{V}$
and $\mathcal{P}$ are the function spaces for velocity and pressure,
respectively.

\subsubsection{Solid phase}
The elasticity problem is resolved similarly as the fluid phase using
P$_1$ finite elements for the displacement field. The stress in the fluid is
transferred to the boundary of the solid phase. The pressure field is
given as nodal values, due to the use of first-order Lagrange
elements, and can therefore be transferred directly to the solid
mesh. However, the viscous stress is a derivative of the velocity
field, and therefore exists as constant values on each
element. Therefore, by stress reconstruction, the stress is
interpolated on the boundary nodes, yielding an error of the order of
the element size. To minimize this error, the mesh was refined near
the fluid-solid boundaries. Additionally, the magnitude of the
viscous stress is orders of magnitude smaller than that of the
pressure, as shall be demonstrated in the next section, yielding an
even smaller relative error in the boundary stress.

The weak problem formulation can be put as follows: Find $\mathbf u'$ in
$\mathcal{V}$ such that
\begin{linenomath*}
\begin{multline}
 \int_{\Omega_\textrm{s}} \ensuremath{\mbox{\boldmath $ \sigma$}}[\mathbf u] : \ensuremath{\mbox{\boldmath $ \epsilon $}}[\mathbf u'] \, \mathrm{d} V
 \\= - \int_{\Gamma_\mathrm{ext} \setminus \Gamma_\textrm{bot}} P_\mathrm{ext} \mathbf n \cdot \mathbf u' \, \mathrm{d} S
 - \int_{\Gamma_\textrm{wall}} P \, \mathbf n \cdot \mathbf u' \, \mathrm{d} S
 + \int_{\Gamma_\textrm{wall}} \mathbf u' \cdot \ensuremath{\mbox{\boldmath $ \sigma$}}_\ell^\mathrm{visc} \cdot \mathbf n \, \mathrm{d} S,
\end{multline}
\end{linenomath*}
for all $\mathbf u' \in \mathcal{V}$, and $\mathbf u(\mathbf x) = \mathbf 0$ for $\mathbf x \in \Gamma_\textrm{wall}$.

\subsection{Probability density functions}
In the simulated samples, the \emph{empirical} probability density
functions (PDFs), $p(\psi)$ for any given scalar field (e.g.~pressure,
stress invariants, fluid velocity components), $\psi$, can be
calculated either on the surface or in the bulk (volume) of the
sample. That is, $p(x) \, \mathrm{d} W$ gives the probability
of finding the value $x$ in an arbitrary infinitesimal volume
{or area} $\mathrm{d} W$.

For the volumetric probability distribution functions, optimal
representation is achieved by weighting each nodal value by the size
of its surrounding volume, similar to its Voronoi cell. For a given
node $i$, this weight can be expressed as
\begin{linenomath*}
\begin{equation}
 w_i = \frac{1}{4 V}\sum_{j \in \mathcal{E}_i } V_j.
\end{equation}
\end{linenomath*}
Here, $\mathcal{E}_i$ is defined as the set of all mesh elements which
contain node $i$, $V_j$ is the volume of element $j$, and $V$ is the
total volume. Similarly, for the surface PDFs, the nodal weight is
found by
\begin{linenomath*}
\begin{equation}
 w_i = \frac{1}{3 A}\sum_{j \in \mathcal{F}_i} A_j
\end{equation}
\end{linenomath*}
where $\mathcal{F}_i$ is defined as the set of all mesh facets which
have node $i$ as a vertex, $A_j$ is the area of facet $j$, and $A$ is
the total area. The PDF is then calculated by normalizing the weighted
histogram of the given field. In order to minimize the effect of
application of external loading, the nodes closest (within 2\%) to the
cubic bounding box are omitted.

\section{Results}
\label{sec:results}
This section presents the results from the coupled fluid-solid
simulations. In turn, we present the results from the fluid, and then
the stress calculations in the solid due to fluid flow and porosity
increase.

\subsection{Main assumptions} 
Our results are sensitive to a series of assumptions made, mainly
related to the discretization of flow in the porous samples:
\begin{itemize}
\item The segmentation process of solid and fluid does not
 unambiguously capture microporosity as some voxels could contain a
 fraction of solid and a fraction of porosity.
\item The removal of disconnected pores and the micropores smaller
 than the voxel size should contribute to stress
 heterogeneities. Note that the removed disconnected pores
 represented a fraction less than 0.05 of the total pore volume.
\item The meshing of the complex microstructure could be done in
  different ways. Note that we are here {using}
  unstructured meshes which better
  approximate the true microstructure{s} than what
  would using e.g.~a cartesian grid.
\item The elastic parameters of the solid phase are assumed to be
 constant throughout the sample.
\item The boundary conditions could have been chosen differently
 (e.g.,~strain controlled rather than stress controlled).
\item The sample has a finite size, limiting the range of length
 scales for the observed spatial correlations.
\end{itemize}
As such, perfect agreement in comparison to experiments should not be
expected. However, the meshes corresponding to snapshots of the sample
at different stages of dissolution are prepared in the same way, and
therefore the \emph{evolution} of the distributions should hold as
long as we consider viscous flow and linear elastostatics. Moreover,
the largest stress concentrations are expected to be found near the
biggest pores, meaning that the discretization is justified, although
e.g.~the ``mesh porosity'' is not the true porosity. {Moreover,
  the fluid-solid solver was validated against cases where
  analytical expressions where available, e.g.~for fluid flow in a
  cylindrical pipe and the stress field around a fluid-filled
  spherical pore. However, as the methods are rather standard and the
  framework is tested by the group of developers, we believe that the
  main sources of error lie in the points above, not in the solver
  itself.}

\subsection{Fluid flow in the pore space}
Here we present the results from pure fluid flow simulations. Due to
the invariance under a linear transformation described in
Sec.~\ref{sec:fluideqs}, the results are given in scaled
units. {Similarly as in section \ref{subsec:fluidphaseweak}, we
  have omitted the corresponding tildes for scaled units.} Physical
values are found by using
Eq{s}.~\eqref{eq:scalefluidvars} and \eqref{eq:unscalefluidstress}.

\begin{figure}[htb]
 \centering
 \includegraphics[width=0.7\textwidth]{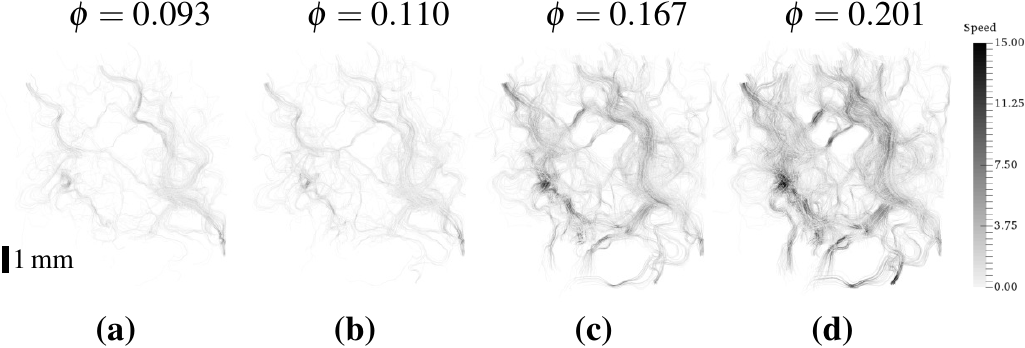}
 \caption{\label{fig:flow}Streamlines of simulated velocity field in
 the sample at different steps of dissolution. Sub-figures
 \textbf{(a)}--\textbf{(b)} correspond to those in
 Fig.~\ref{fig:mesh_fluid}.}
\end{figure}

In Fig.~\ref{fig:flow}, the simulated flow field is visualized by
streamlines, i.e.~integrated Lagrangian trajectories of the velocity
field. As expected with increasing porosity, the flow through the
sample increases. At low porosity, a few preferential flow paths are
present. At higher porosities, more paths appear and cross-link with
each other.

\begin{figure}[htb]
 \centering
 \includegraphics[width=0.99\textwidth]{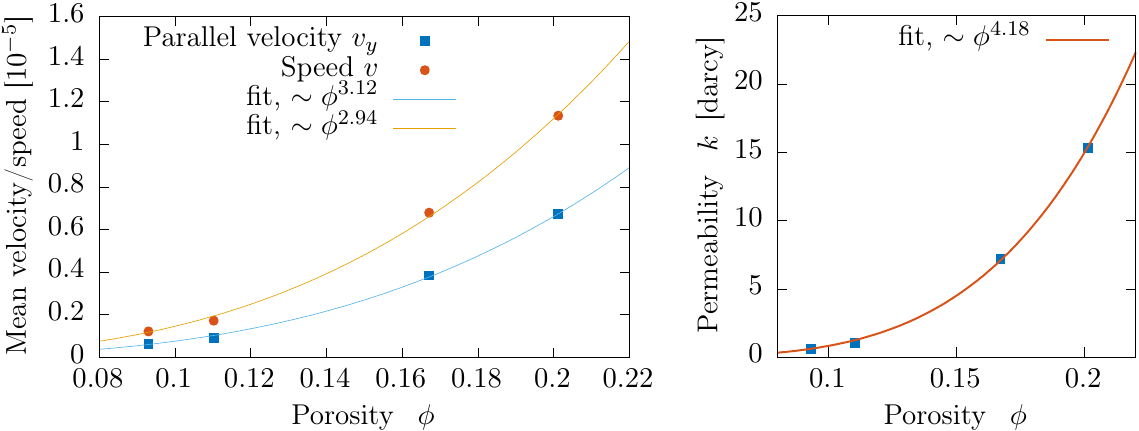}
 \caption{\label{fig:speed_vs_porosity} {Left:} Mean axial
   velocity/speed in the bulk of the fluid versus porosity in the
   sample{, in scaled units}. {Right: Corresponding permeability in SI units.}}
\end{figure}

{Flow through porous media is on the \emph{macroscale} governed
  by Darcy's law,}
\begin{equation}
  {\mathbf q = - \frac{k}{\mu} \ensuremath{\mbox{\boldmath$ \nabla $}}{P}},
\end{equation}
{where $\mathbf q$ is the flux (discharge per area) and $k$ is the
  permeability. Here, the flux is related to the mean velocity through
  the relation $\mathbf q = \phi \mathbf v$. Historically, much effort has been
  devoted to relating the permeability $k$ to porosity $\phi$, the
  most popular being the Kozeny--Carman relation commonly expressed as
  $k = C \phi^3 / (1-\phi)^2$
  \citep{costa2006permeability,matyka2008tortuosity}, where $C$ is a
  constant of dimension $(\textrm{length})^2$ related to the geometry
  of the porous medium.}  The mean absolute velocity
  (speed) and the mean axial velocity (parallel to
the pressure gradient), are plotted as functions of porosity,
$\left< v_y \right> (\phi)$, in {the left panel of}
Fig.~\ref{fig:speed_vs_porosity}, and display superlinear increase with porosity. The solid lines
represent power-law fittings to the data, which both
yield exponents $\simeq 3$.  Taking the pressure drop to be constant,
the Kozeny--Carman relation predicts that the flux {through any
  cross section of the sample}
$q { = \mathbf q \cdot \hat{\mathbf n} = \phi \left< v_y \right>}$
of a porous media should depend on porosity as
\begin{linenomath*}
\begin{equation}
 q \propto \frac{\phi^3}{(1-\phi)^2}.
\end{equation}
\end{linenomath*}
This means that the average velocity should scale as
$\left< v \right> \sim q/\phi \sim \phi^2/(1-\phi)^2$. The
fittings shown in Fig.~\ref{fig:speed_vs_porosity}
are thus not in quantitative agreement with Kozeny--Carman
relations. However, this is not unexpected, as Kozeny--Carman
relations, being derived for packed beds, are usually more applicable
to configurations such as high porosity sandstone, and less so for
low-porosity limestone {undergoing dissolution}. {This
  observation is consistent with the data presented in the right panel
  of Fig.~\ref{fig:speed_vs_porosity}, where permeability $k$ in
  physical units is plotted as a function of porosity $\phi$, and the
  relationship $k \sim \phi^4$ grows faster than the prediction from
  the Kozeny--Carman relation. The behaviour seen here is comparable
  to the data reported by \citet{ehrenberg2006porosity}, although the
  magnitude is somewhat higher here, especially as dissolution
  progresses. The permeability calculated here, however, coincides
  well with the permeability reported in the original experiment
  \citep{noiriel2004}.}

\begin{figure}[htb]
 \centering
 \includegraphics[width=0.6\textwidth]{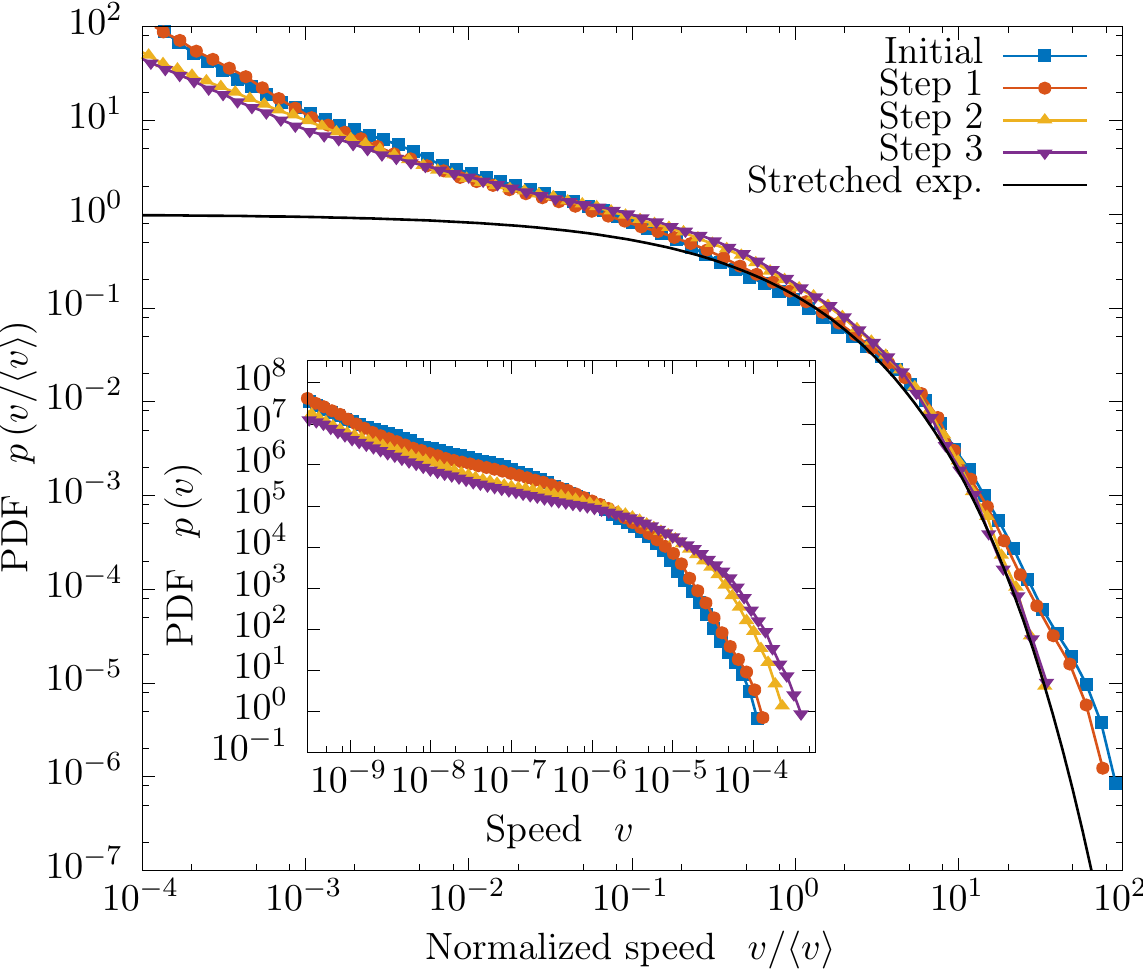}
 \caption{\label{fig:pdf_speed}Probability density of the fluid
 velocity in the pore space at different steps of dissolution. The
 velocity distributions are collapsed upon rescaling by the average
 velocity. Inset: Non-normalized distribution.}
\end{figure}

Fig.~\ref{fig:pdf_speed} displays the measured probability
distribution of fluid speed in the four volumes. The inset of
Fig.~\ref{fig:pdf_speed} shows the raw (non-normalized) data, with a
shift of the distribution towards higher speed as porosity is
increased. The relevant features of the distributions are extracted by
rescaling the speed, $v$, by the mean speed $\left< v \right>$ in each sample
(displayed in Fig.~\ref{fig:speed_vs_porosity}), as shown in the main
panel of Fig.~\ref{fig:pdf_speed}. The distributions collapse, apart
from in the tail (possibly due to the finite size of the computational
meshes). As the porosity increases, the distribution approaches a
stretched exponential function for large $v$,
\begin{linenomath*}
\begin{equation}
 p \left( \hat v \right) \sim \exp\left( - \alpha \hat v ^\beta \right) 
\end{equation}
\end{linenomath*}
where $\hat v = v /\left< v \right>$, the exponent $\beta \simeq 1/2$, and the
scale parameter $\alpha\simeq0.25$. This observation is consistent
with established velocity statistics for disordered porous media
\citep{matyka2016} and the stretched exponential distribution can
theoretically be inferred by considering the porous media as a
collection of cylinders with exponentially distributed radii
\citep{holzner2015intermittent}.

\begin{figure}[htb]
 \centering
 \includegraphics[width=0.6\textwidth]{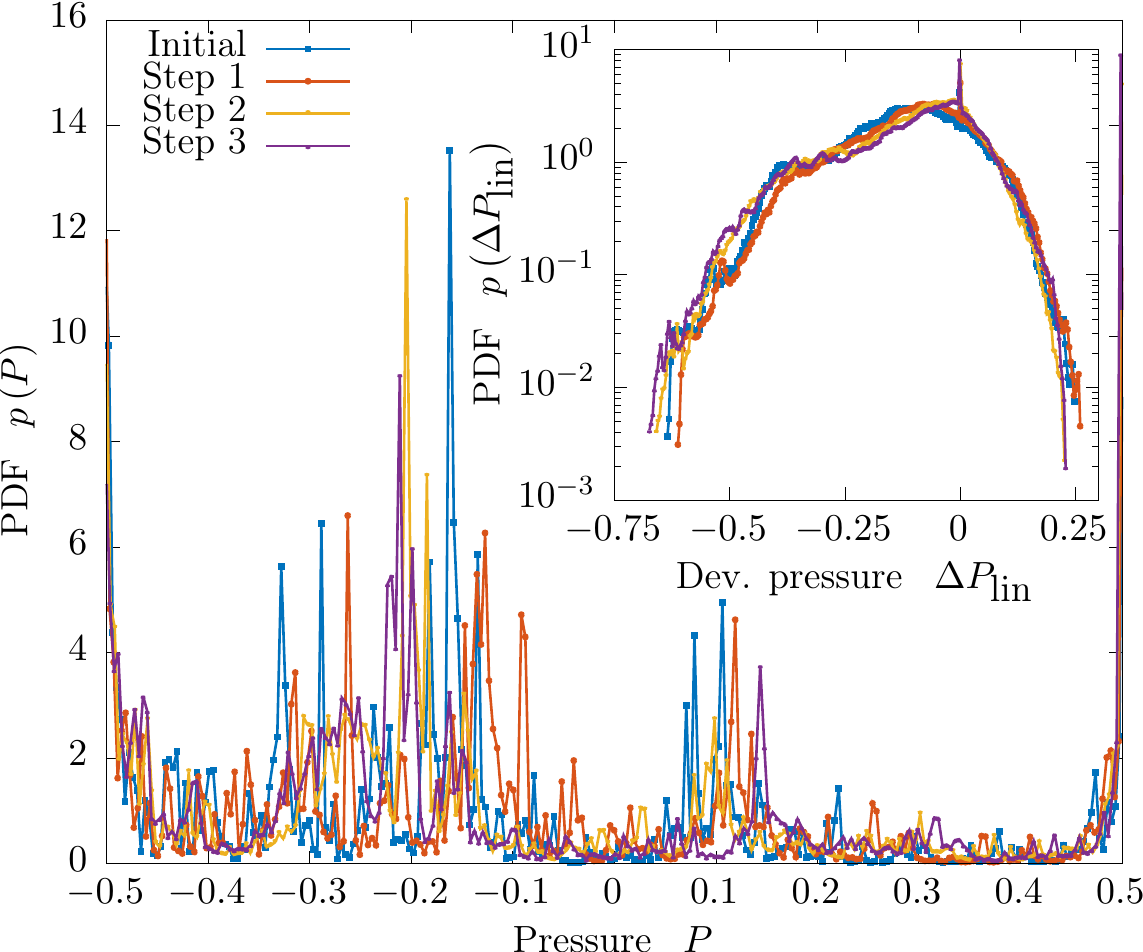}
 \caption{\label{fig:pdf_pressure}Fluid pressure statistics at the
 pore walls in the sample at different steps of dissolution.
 }
\end{figure}

The fluid pressure distribution is shown in Fig.~\ref{fig:pdf_pressure},
and displays a highly heterogeneous distribution between the inlet
pressure $P_\textrm{in}=1/2$ and the outlet pressure $P_\textrm{out}=-1/2$. The
distribution is characterized by fluctuations (spikes) that are
interpreted to be related to a heterogeneous distribution of dead-end
pores. To quantify the heterogeneity in the pressure field arising
from the flow, the deviation from a linear pressure profile (like what
appears in hydrostatics with a constant gravitational force) can be
calculated as:
\begin{linenomath*}
\begin{equation}
 \Delta P_\textrm{lin} (\mathbf x) = P (\mathbf x) - P_\textrm{lin} (\mathbf x),
\end{equation}
\end{linenomath*}
where $P_\textrm{lin} (\mathbf x) = 1/2 - y$, $y \in [0, 1]$ is the
scaled coordinate along the direction of the imposed pressure drop,
such that $y=0$ corresponds to the inlet face, and conversely for
$y=1$. The resulting distribution is shown in the inset of
Fig.~\ref{fig:pdf_pressure}. The distribution is seen to be sharply peaked
around $\Delta P_\textrm{lin} = 0$, due to the fixed pressure at
inlet and outlet, but apart from this, slightly skewed towards
negative values. This indicates a geometrical asymmetry in the sample:
more dead-end pores stretch from the top (low pressure) to the bottom
of the samples, than the other way around.

\begin{figure}[htb]
 \centering
 \includegraphics[width=0.4\textwidth]{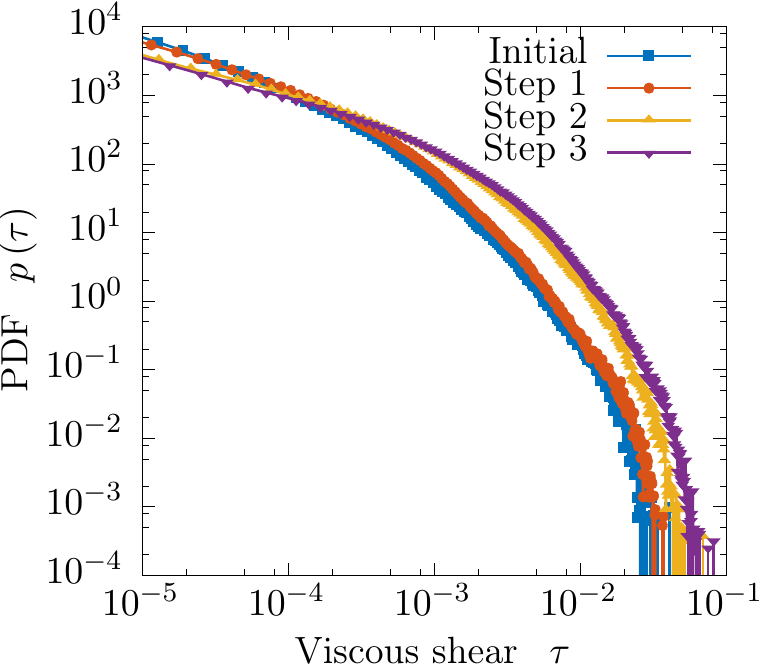}
 \caption{\label{fig:shear_statistics}Fluid shear stress statistics
 at the pore walls in the sample at different steps of
 dissolution.}
\end{figure}

The fluid \emph{pressure} exerts a \emph{normal} traction $p \hat{\mathbf{n}}$
upon the solid matrix. The viscous flow field, on the other hand,
contributes to a \emph{tangential} traction. The \emph{scaled} viscous
stress tensor is given by
\begin{linenomath*}
\begin{equation}
 \ensuremath{\mbox{\boldmath $ \sigma$}}_\mathrm{visc} = \ensuremath{\mbox{\boldmath$ \nabla $}} \mathbf v + \ensuremath{\mbox{\boldmath$ \nabla $}} \mathbf v^\top,
\end{equation}
\end{linenomath*}
and to quantify this (while suppressing the influence of the boundary
normal, which must be reconstructed on nodes, introducing an error of
order element size) we report the distribution of $\tau$, the largest
absolute eigenvalue of $\sigma_\mathrm{visc}$; sampled over the surface nodes
(as in \citep{voronov2010}). The resulting distribution is shown in
Fig.~\ref{fig:shear_statistics}. From the latter figure, it is clear
that the viscous forces are, generally, at least 1--2 orders of
magnitude lower than the pressure drop, which again is lower than the
base pressure $P_0$ and the external pressure $P_\mathrm{ext}$. We emphasize
that this observation is \emph{independent} of the value of the
viscosity $\mu$, as the equations are linear and
therefore only one unique solution for the stress field exists apart
from a scaling (by $\Delta P$) and a shift (by $P_0$).

\subsection{Stress in the porous solid}
In the following, the results from calculating the state of stress in
the solid due to fluid flow and porosity increase are reported. {In all cases, a Poisson's ratio $\nu = 0.3$, an external pressure $P_\mathrm{ext}=\SI{2.2E+7}{\pascal}$, and a base pressure $P_0=\SI{1.0E+7}{\pascal}$ were used.}

\subsubsection{Measures of state of stress in the solid}
In order to assess the impact of applied external stress on the porous
sample, we consider frame-invariant quantities. Combinations of the
first two invariants of the stress tensor will therefore be used
($I_1$ and $I_2$). First, the \emph{mechanical pressure} is defined as
\begin{linenomath*}
\begin{equation}
 P_\mathrm{mech} = - \frac{\mathrm{tr}{ {(} \ensuremath{\mbox{\boldmath $ \sigma$}} {)}}}{3}.
\end{equation}
\end{linenomath*}
Secondly, the \emph{von Mises stress} \citep{mises1913} is defined by
\begin{linenomath*}
\begin{equation}
 \sigma_\textrm{vM} = \sqrt{ \frac{3}{2} \ensuremath{\mbox{\boldmath $ \sigma$}}_\textrm{dev} : \ensuremath{\mbox{\boldmath $ \sigma$}}_\textrm{dev} },
\end{equation}
\end{linenomath*}
where the deviatoric stress tensor is defined by
$\ensuremath{\mbox{\boldmath $ \sigma$}}_\textrm{dev} = \ensuremath{\mbox{\boldmath $ \sigma$}} + P_\mathrm{mech} \mathbf I$. The von Mises stress
is commonly used to predict yielding of materials under multi-axial
loading. The tightly related von Mises \emph{yield criterion} states
that a material starts to deform irreversibly when $\sigma_\textrm{vM}$
reaches a certain critical threshold $\sigma_\textrm{yield}$. As such, it is a
measure of how close to fracturing the material is when considering a
brittle material such as a rock in the first kilometers of the Earth's
crust. We note that we could as well have considered some of the
common alternatives to the von Mises stress, such as the maximum
principal stress (strain) i.e.~the largest eigenvalue of the stress
(strain) tensor, $\sigma_1$ (or
$\sigma_1 - \nu (\sigma_2 + \sigma_3)$), and the results are largely
similar.

\subsubsection{Influence of pressure drop over fluid on stress in the solid}
\begin{figure}[htb]
 \centering
 \includegraphics[width=0.6\textwidth]{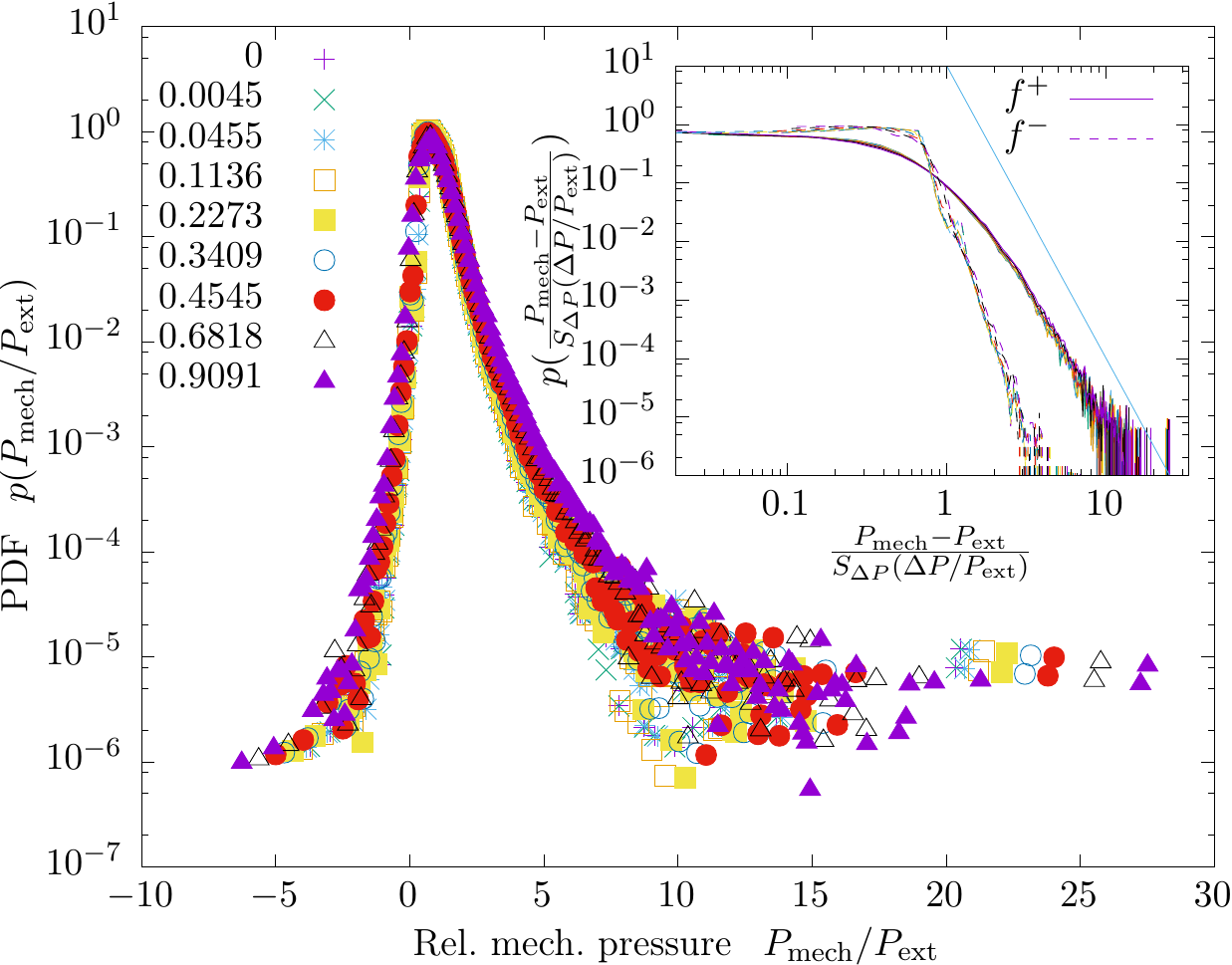}
 \caption{\label{fig:pdf_p_mech_init}Probability density
 function of the mechanical pressure $P_\mathrm{mech}$ at the pore walls in
 the sample before dissolution, at various imposed fluid pressure drops
 $\Delta P$. Here, $P_\mathrm{ext}=\SI{2.2E+7}{\pascal}$,
 $P_0=\SI{1.0E+7}{\pascal}$. Inset: data collapse by
 rescaling. The straight line shows power-law decay
 $p(x) \sim x^{-5}$ as a guide to the eye. Similar plots are also found for the various dissolution steps.}
\end{figure}

Fig.~\ref{fig:pdf_p_mech_init} shows the probability distributions of
$P_\mathrm{mech}$ in the sample before dissolution, normalized by the external
pressure $P_\mathrm{ext}$, obtained for various fluid pressure drops
$\Delta P \in [0, P_\mathrm{ext}]$. The distributions are peaked around
$P_\mathrm{mech}/P_\mathrm{ext}=1$, with a markedly heavy tail for large $P_\mathrm{mech}$.

In the inset of Fig.~\ref{fig:pdf_p_mech_init} the distributions are seen
to collapse by the normalization
\begin{linenomath*}
\begin{equation}
 \hat P = \frac{P_\mathrm{mech} - P_\mathrm{ext}}{P_\mathrm{ext} S_{\Delta P} (\frac{\Delta P}{P_\mathrm{ext}} )}.
\end{equation}
\end{linenomath*}
Here, $S_{\Delta P} (\Delta P /P_\mathrm{ext} )$ is a scaling
function described below.

For large $\hat P$, a power-law {behaviour}
$p(\hat P) \sim \hat P^{-\gamma}$, {where}
$\gamma \simeq 5${,} can {possibly} be
observed. The support is {however}
only over one order of magnitude and therefore other distributions
might {also} provide good fits, e.g.~a
stretched exponential distribution.

\begin{figure}[htb]
 \centering
 \includegraphics[width=0.6\textwidth]{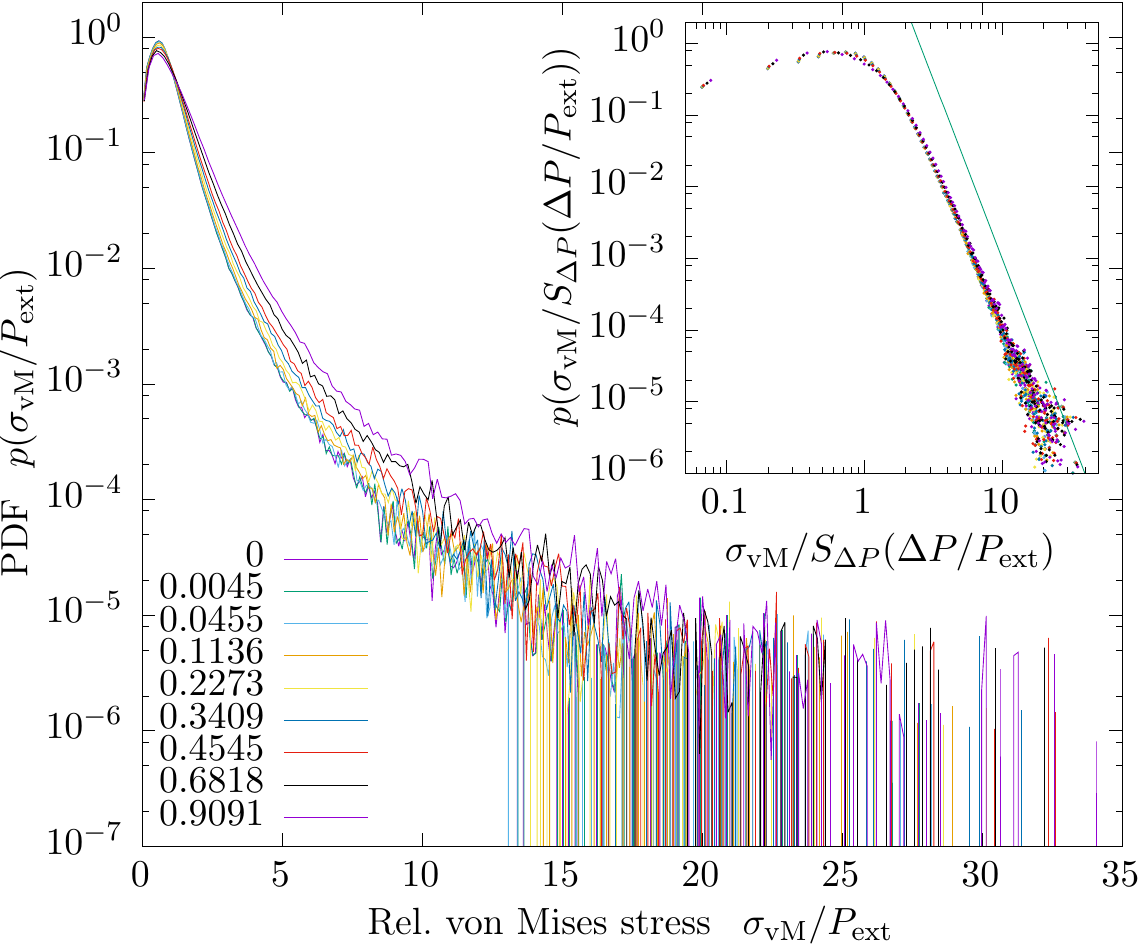}
 \caption{\label{fig:pdf_sigma_vm_init}Probability density function
 of the von Mises stress $\sigma_\textrm{vM}$ at the pore walls in the
 initial step of the sample, at various imposed pressure drops
 $\Delta P$. $P_\mathrm{ext}$ and $P_0$ are the same as in
 Fig.~\ref{fig:pdf_p_mech_init}. Inset: data collapse by rescaling. A
 power-law decay with exponent $-5$ is shown as a guide to the
 eye. Similar plots are also found for the various dissolution steps.}
\end{figure}

The probability distributions of the von Mises stress
$\sigma_\textrm{vM}$, corresponding to Fig.~\ref{fig:pdf_p_mech_init} are
shown in Fig.~\ref{fig:pdf_sigma_vm_init}. The distributions of
$\sigma_\textrm{vM}$ display similar characteristics as the
distributions for $P_\mathrm{mech}$. Scaling by
$S_{\Delta P} ( \Delta P / P_\mathrm{ext} )$ yields the same
data collapse as for $\hat P$, i.e.~distributions of
\begin{linenomath*}
\begin{equation}
 \hat{\sigma}_\textrm{vM} = \frac{\sigma_\textrm{vM}}{P_\mathrm{ext} S_{\Delta P} (\frac{\Delta P}{P_\mathrm{ext}} )} 
\end{equation}
\end{linenomath*}
are independent of $\Delta P$.

\begin{figure}[htb]
 \centering
 \includegraphics[width=0.4\textwidth]{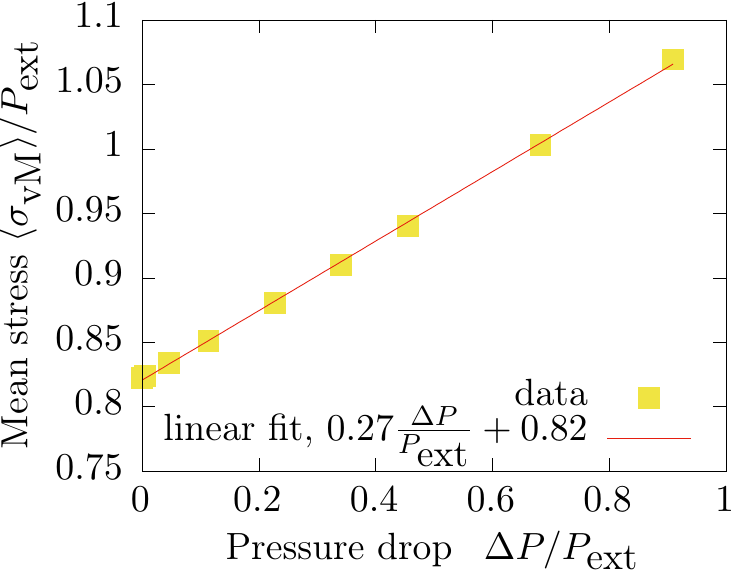}
 \caption{\label{fig:mean_sigma_vm_vs_pressure_drop}Mean von Mises
 stress as a function of the pressure drop $\Delta P$ inducing flow
 through the sample. The means are taken over the
 distributions in Fig.~\ref{fig:pdf_sigma_vm_init}.}
\end{figure}

In Fig.~\ref{fig:mean_sigma_vm_vs_pressure_drop}, the data points show the
mean of the pore-wall distribution of $\sigma_\textrm{vM}/P_\mathrm{ext}$, plotted as a
function of (normalized) pressure drop. A linear least-squares fit
is used to determine the scaling function:
\begin{linenomath*}
\begin{equation}
 S_{\Delta P} ( \frac{\Delta P}{P_\mathrm{ext}} ) = 0.27 \frac{\Delta P}{P_\mathrm{ext}} + 0.82.
 \label{eq:scale_dp}
\end{equation}
\end{linenomath*}

\subsection{Influence of dissolution and increasing porosity}
\begin{figure}[htb]
 \centering
 \includegraphics[width=0.6\textwidth]{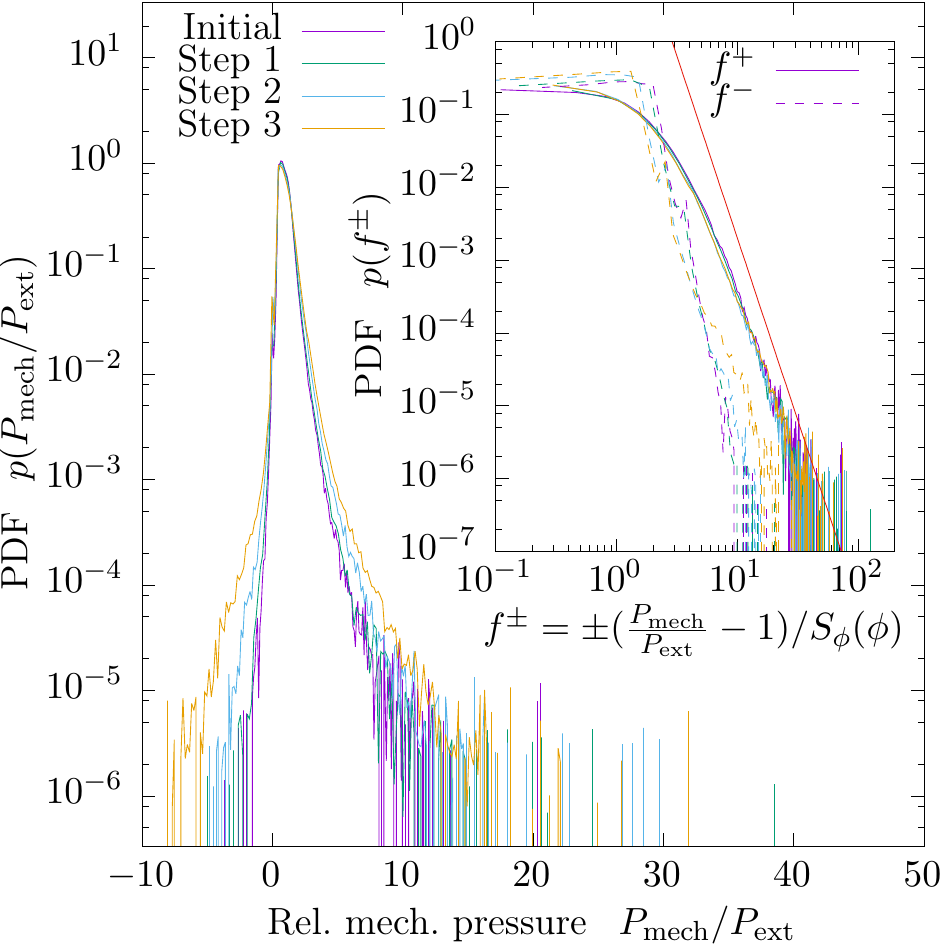}
 \caption{\label{fig:pdf_p_mech_dp-0}Probability density function of
   the mechanical pressure at the pore walls for all dissolution
   steps, at fixed $\Delta P = 0$. $P_\mathrm{ext}$ and $P_0$ are the same
   as in Fig.~\ref{fig:pdf_p_mech_init}. Inset: data collapse by
   rescaling. The line shows a power-law with exponent -5 as a guide
   to the eye. Similar plots exist for other pressure drops.}
\end{figure}

The probability distributions of the mechanical pressure at the pore
walls of the sample at different stages of dissolution are shown in
Fig.~\ref{fig:pdf_p_mech_dp-0}. Here, the pressure values used were
$P_\mathrm{ext}=\SI{2.2E+7}{\pascal}$, $P_0=\SI{1.0E+7}{\pascal}$, and
$\Delta P=0$. The peaks of the distributions are located at
the same position, $P_\mathrm{mech}=P_\mathrm{ext}$, but the distributions become wider as
the porosity is increased, indicating more stress concentration and
more stress shadows with increasing dissolution.

To segregate the distribution{s} of values above ($+$) and below
($-$) the peak at $P_\mathrm{mech} = P_\mathrm{ext}$, we define
\begin{linenomath*}
\begin{equation}
 f^\pm = \pm \frac{P_\mathrm{mech}- P_\mathrm{ext}}{P_\mathrm{ext} S (\frac{\Delta P}{P_\mathrm{ext}}) S_\phi(\phi)}
 \label{eq:fpm}
\end{equation}
\end{linenomath*}
where the scaling function $S_\phi(\phi)$ is defined below (see
Eq.~\ref{eq:scale_phi}). As shown in the inset of
Fig.~\ref{fig:pdf_p_mech_dp-0}, the resulting distributions largely
collapse: the distributions of $f^+$ are seen to fall
onto the same curve, while $f^-$ displays a slightly varying slope with dissolution step.

\begin{figure}[htb]
 \centering
 \includegraphics[width=0.6\textwidth]{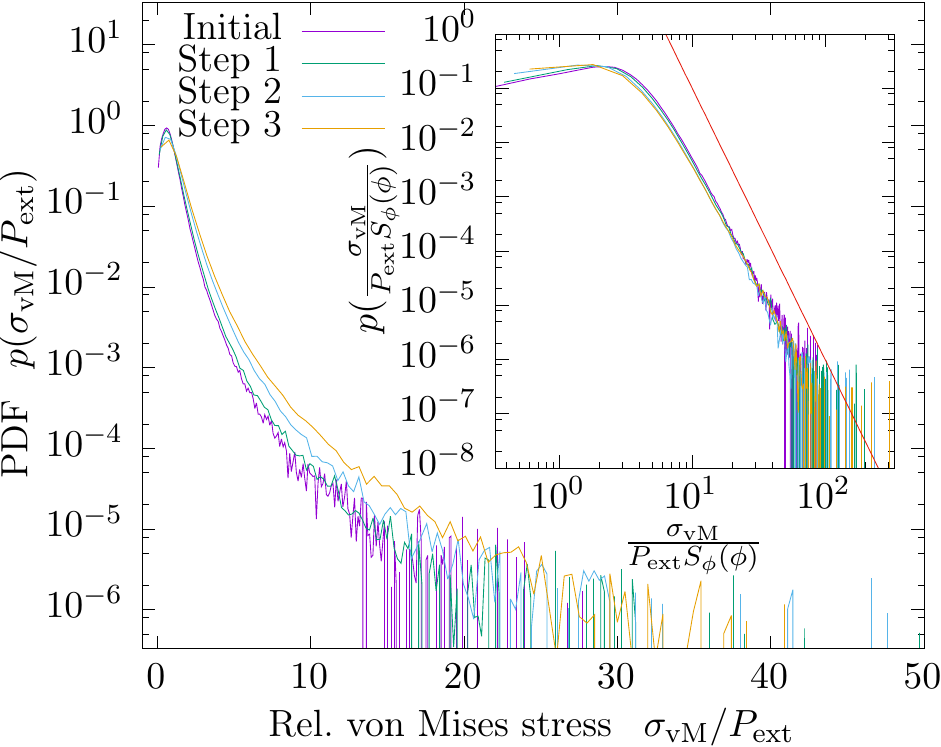}
 \caption{\label{fig:pdf_sigma_vm_p0-0}Probability density function
 of the von Mises stress at the pore walls for all samples, at
 fixed $\Delta P = 0$. $P_\mathrm{ext}$ and $P_0$ are the same as in
 Fig.~\ref{fig:pdf_p_mech_init}. Inset: data collapse by
 rescaling. Similar plots exist for other pressure
 drops.}
\end{figure}

\begin{figure}[htb]
 \centering
 \includegraphics[width=0.6\textwidth]{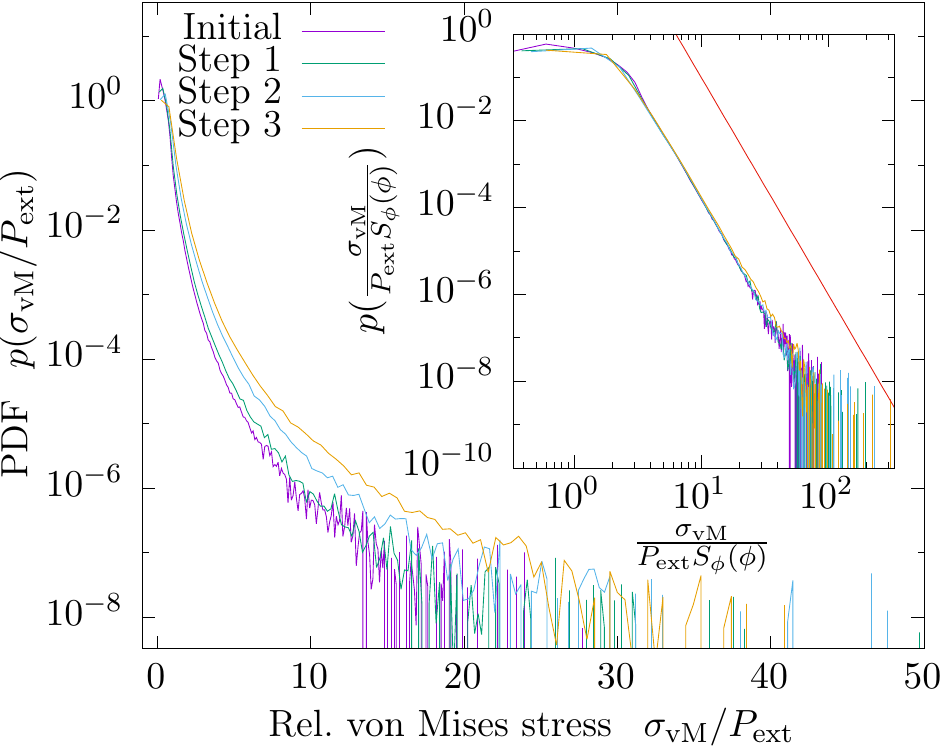}
 \caption{\label{fig:pdf_sigma_vm_p0-0_bulk}Probability density
   function of the von Mises stress in the bulk for all samples, at
   fixed $\Delta P = 0$. $P_\mathrm{ext}$ and $P_0$ are the same as in
   Fig.~\ref{fig:pdf_p_mech_init}. Inset: data collapse by rescaling,
   and a superimposed power law with exponent -5 as a guide to the
   eye. Similar plots exist for the other pressure drops.}
\end{figure}

The probability distribution of the (relative) von Mises stress is
plotted for the pore walls in Fig.~\ref{fig:pdf_sigma_vm_p0-0} and for the
solid bulk in Fig.~\ref{fig:pdf_sigma_vm_p0-0_bulk}. The pore-wall
distributions display the same behavior and collapse by $S_\phi(\phi)$
as that of $f^+$ above. In comparison, the bulk distribution extends
the suggested power-law distribution,
$p(\sigma_\textrm{vM}) \sim \sigma_\textrm{vM}^{-\gamma}$, $\gamma \simeq 5$, for large
$\sigma_\textrm{vM}$.


\begin{figure}[htb]
 \centering
 \includegraphics[width=0.4\textwidth]{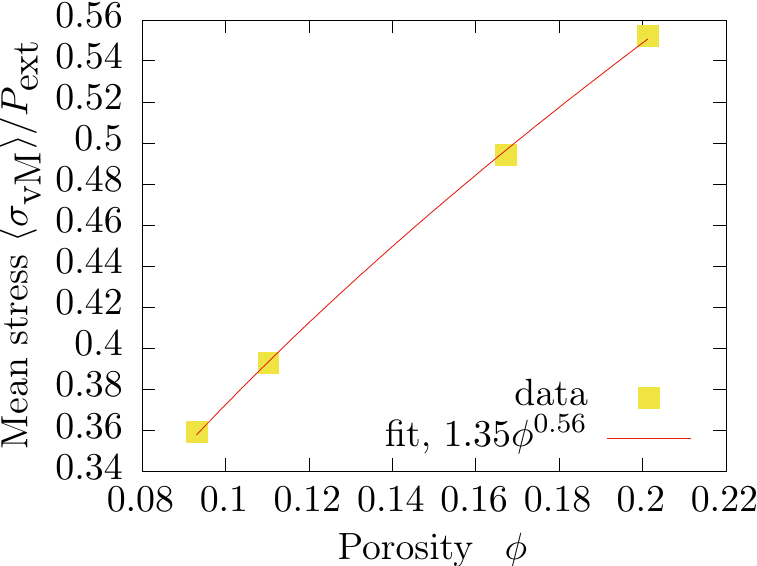}
 \caption{\label{fig:mean_sigma_vm_vs_porosity}Bulk mean von Mises
 stress versus porosity. The data points correspond to the
 distributions in Fig.~\ref{fig:pdf_sigma_vm_p0-0_bulk}.}
\end{figure}

The bulk averages of $\sigma_\textrm{vM}$, corresponding to the probability
distributions shown in Fig.~\ref{fig:pdf_sigma_vm_p0-0_bulk}, are plotted
in Fig.~\ref{fig:mean_sigma_vm_vs_porosity}. The scaling function
$S_\phi(\phi)$ is approximated as a fit to these points. We
expect no deviatoric stress at $S_\phi(0) = 0$, and the simplest form
satisfying this is
\begin{linenomath*}
\begin{equation}
 S_\phi(\phi) = C \phi^\beta.
 \label{eq:scale_phi}
\end{equation}
\end{linenomath*}
Here, the exponent $\beta \simeq 0.56$ yields the
best fit of the experimental data
using a least-square method. An even better fit would be achieved by
using more complicated expressions with more fitting parameters, but
for that to be justified one would also have required more than the
four porosity levels available herein. Alternatively, exponential fits
could be used, analogous to the compiled data of critical axial stress as a
function of porosity in limestones summarized in \citep[Sec.~3.1.3]{croize2013}.

\subsection{Common probability density functions}
As a consequence of the above analysis, all distributions considered
will collapse onto the same master curves, by the scaling relationships
\begin{linenomath*}
\begin{equation}
 \hat{\tilde{P}} = \frac{P_\mathrm{mech} - P_\mathrm{ext}}{P_\mathrm{ext} S_{\Delta P}(\frac{\Delta P}{P_\mathrm{ext}}) S_\phi(\phi)},
\end{equation}
\end{linenomath*}
and
\begin{linenomath*}
\begin{equation}
 \tilde{\sigma}_\textrm{vM} = \frac{\sigma_\textrm{vM}}{P_\mathrm{ext} S_{\Delta P}\left(\frac{\Delta P}{P_\mathrm{ext}}\right) S_\phi (\phi) },
\end{equation}
\end{linenomath*}
where the scaling functions
$S_{\Delta P} (\Delta P/P_\mathrm{ext})$ and $S_\phi(\phi)$ are
given by Eqs.~\eqref{eq:scale_phi} and \eqref{eq:scale_dp}, respectively. This unified
description of stress heterogeneities in the limestone sample studied
here represents the main outcome of the present study.

\section{Discussion}\label{sec:discussion}
\subsection{Complexity of fluid flow in a porous medium}
\citet{brown1987} solved Reynolds equations in 2D in a synthetic rough
aperture fracture and showed that for low aperture, the roughness of
the walls had a significant effect, leading to flow channeling. In a
porous medium, the pore structure complexity generates a wide range of
flow velocities, from the fast advective flows in the main channels, to
the very slow diffusive flows in dead ends where the fluid is almost
stagnant and rarely mix with that in the main channels
\citep{bijeljic2013}. The pore heterogeneities have a strong effect on
the long-range spatial correlations of the flux. Numerical simulations
show that the distributions of the kinetic energy and the velocity in
the fluid follow power laws over at least five orders of magnitude
\citep{andrade1997,makse2000} and that the flow is correlated in space
and time \citep{leborgne2008}, leading to intermittency
\citep{deanna2013}. Increasing the complexity of pore geometry, from a
simple bead-pack porous medium to natural rock samples with micropores
and microfractures increases as well the range of velocities observed
in the fluid. The velocity distribution is characterized by a main
peak, controlled by the pressure drop imposed on the system, and a
tail of slow velocities that increases with pore network complexity
\citep{bijeljic2013,jin2016}. Based on Lattice Boltzmann simulations,
\citet{matyka2016} proposed that the probability distribution function
of fluid velocity, for velocities larger than the average fluid
velocity, follows a ``power-exponential'' law. This is in contrast with
other studies which have proposed either a Gaussian or an exponential
distribution \citep{mansfield1996,datta2013,bijeljic2013,lebon1996}.

With regards to the speed distributions presented in
Sec.~\ref{sec:results}, a stretched exponential probability density
function provides a good fit for large speeds. A shifted, stretched
exponential (``power-exponential'') distribution, as proposed by
\citet{matyka2016}, would also be in agreement with our results, but
{this} would require introducing another fitting parameter. Moreover, the
evolving pore structure due to dissolution in our sample does not
significantly alter the functional dependence of the probability
density function of fluid velocities, when rescaled by the average
velocity.

\subsection{Coupling fluid flow and deformation}
The fluid flow in the porous medium exerts both shear and normal
stress on the solid walls, as shown numerically for a rough fracture
\citep{lo2014}. Because of the complexity of the porous medium,
additional complexity of the flow pattern exists and long range
correlations in the stress distribution at the solid interface
emerge.

Flow-induced stresses have been modeled for several biological
applications where porosity of the medium was quite high (above 80 \%)
and the solids were very soft. Under these conditions, numerical
simulations indicate that the fluid viscous stress at the pore walls
follows a gamma distribution \citep{voronov2010}. Numerical models of
fluid flow in highly deformable elastic porous media indicated that,
as the elastic solid deforms under flow, the relationship between
pressure drop and flux becomes non-linear and saturates for large
pressure gradients \citep{hewitt2016}. Hysteresis due to the coupling
between fluid and solid can emerge \citep{Guyer2015}. \citet{pham2014}
calculated the stress exerted by a fluid around a spherical solid
using Lattice Boltzmann simulations, and the existence of areas with
stress concentration on the solid, and log-normal stress distribution
was observed. However, in all these studies, the porosity was quite
large and/or the solids were very soft and relevant for bioengineering
applications. This renders comparison with solids that are stronger
and with lower porosity, such as rocks, challenging.

In rocks, elastic deformations are quite small, usually below one
percent, before irreversible strain occurs. Depending on stress and
the mechanisms of irreversible deformation, such as closure and opening
of microcracks or pore collapse, the relationship between porosity and
permeability evolves, controlling the pore pressure gradient
\citep{david1994}. Under loading, the microscale heterogeneities
control both the initiation of microcracks and the overall strength of
the material. Using a 2D discrete element modeling approach applied
to a granite rock, \citet{lan2010} showed a difference between
geometrical heterogeneities (i.e.~difference of grain size), which
control the nucleation of microfractures and initiation of damage, and
strength heterogeneities at the grain contacts (i.e.~elastic
stiffness), which control the overall strength of the solid under
uniaxial loading. In these simulations, the stresses inside the grains
show a normal distribution for both the maximum and the minimum
principal stress, with an average value which corresponds to the
external loading. Conversely, the normal stress at grain contacts
shows a bimodal distribution. Some contacts are under tensile normal
stress conditions and provide sites for the nucleation of extensional
microfractures.

The evolution of elastic parameters and permeability during small
elastic deformations of a Bentheim sandstone was experimentally
measured and successfully modeled using X-ray microtomography images
where unstructured meshes were built \citep{jasinski2015}. The effect
of fluid viscosity on the effective elastic properties of rocks and
the attenuation of elastic waves was studied in \citep{saenger2011} by
solving the dynamic elastic equation in 3D rock samples imaged with
X-ray microtomography. {Other researchers have simulated deformation
  of calcium carbonates with a back-coupling to flow through
  dissolution \citep{nunes2016} and precipitation \citep{jiang2014},
  although without accounting for the stress distribution in the solid
  matrix. In the present work, the fluid-solid coupling is only
  one-way and therefore the effect of flow on changing pore-space
  geometry can not be assessed.}

By considering probability density functions of bulk and pore-wall
properties, the results presented in Sec.~\ref{sec:results} show that
for a steadily flowing fluid in the pore space of a limestone, the
dominating force from the fluid stems from the base pressure in the
solid, as the viscous force generated by the fluid is generally orders
of magnitude lower. This implies that under such conditions, the
viscous stress is of minor importance. Moreover, the stress
distributions are controlled by the pressure drop $\Delta P$
in a simple manner. In particular, the position of the tail of the
distributions of stress in the sample may ultimately depend on the
maximum difference between external and internal pressure. This
{broad} tail, with a power-law decay
with a quite strong exponent of -5{,} has the following
consequence: a slight increase in fluid pressure or a small amount
{of} dissolution will significantly
{increase} the number of locations in the solid where the von
Mises criteria (or another failure criteria) will be
reached. A consequence of such behavior is the
following. It is known that the injection or removal of fluid at depth
can to trigger induced seismicity \citep{talwani1984}. Recent field
observations at the outcrop scale show that a small fluid injection
can trigger microearthquakes at some distance from the injection point
\citep{guglielmi2015}. If they can be extended to other kinds of
rocks, our results, with a heavy power-law tail of stress
heterogeneities, show that a small change in fluid pressure can drive
a significant volume of the rock towards failure. The nature of
microstructural heterogeneities and their relationships to fluid flow
and stress would then provide an additional explanation of induced
seismicity.

{Whether the observed self-similarity persists if the porosity
  is increased beyond the range considered here, is an open question,
  and could be assessed e.g.~by using tomography data from experiments
  where more dissolution is performed. However, in the Earth's crust
  failure would occur before reaching a high porosity, which is what
  happens for example in karst with the formation of caves. Further,
  how the distribution changes if such failure occurs, i.e.~the stress
  heterogeneity leads to fractures, is an interesting point in
  question. We expect the self-similar behaviour will reach its end at
  latest when the first failure occurs, as the solid matrix will then
  reorganize itself.}

\section{Conclusion}\label{sec:conclusion}
We have in this work computationally studied how an evolving
microstructure influences fluid flow in the pore space of a rock, and
how fluid flow influences the state of stress in the solid phase. We
have considered a limestone which has been scanned at four stages of
dissolution using X-ray microtomography.

Steady incompressible laminar fluid flow in the sample at each stage
of dissolution was computed by solving Stokes' equation{s}. By
assuming negligible displacement of the {fluid-solid} boundary
due to elastic deformation, the stress field from the fluid enters as
a boundary condition on the solid, yielding a one-way numerical
coupling. Both the fluid and the solid problems were solved
numerically using the finite element method through the FEniCS/DOLFIN
framework.

Our main finding is that, as the rock is dissolved, and as the
pressure drop driving the fluid flow is increased, the distribution of
heterogeneous stress in the sample evolves in a self-similar manner.
In particular, the probability distributions of the mechanical
pressure and the von Mises stress can be collapsed onto the same curve
by a normalization. The common master curves display a broad
distribution, with a suggested power-law tail for high stresses. The
broad tail shows that the rock is very sensitive to small
perturbations, and a slight fluid pressure increase locally would
drive a significant number of local heterogeneities toward failure. We
propose that this heavy tail can be used as a simple criterion for the
integrity of porous rocks.

Whether the observed self-similar evolution is restricted to
dissolution processes remains to be answered. For example, do other
morphology-changing processes, such as fracturing or precipitation
(lowering porosity) evolve similarly? A more fundamental question is
related to identifying the link between pore geometry and the velocity
distribution/stress distribution. Future work will include a
back-coupling from solid deformation to fluid flow, yielding transient
dynamics of fracture and/or precipitation--dissolution processes.

\begin{acknowledgments}
  The authors thank Catherine Noiriel, University of Paul Sabatier,
  Toulouse, for generously providing the segmented voxelated data of
  the porous sample. Stimulating discussions with Noiriel, Frans Aben
  and Luiza Angheluta were greatly appreciated. This project has
  received funding from the European Union's Horizon 2020 research and
  innovation programme through Marie Curie initial training networks
  under grant agreements No.~642976 (ITN NanoHeal) and 316889 (ITN
  FlowTrans). The project further received funding from the Villum
  Foundation through the grant ``Earth Patterns'' and the Norwegian
  Research Council through the HADES project. The data supporting this
  paper are available by contacting the corresponding author at
  \texttt{linga@nbi.dk}.
\end{acknowledgments}

\end{document}